\documentclass[preprint,amsmath,amssymb,aps]{revtex4-1}
\usepackage{graphicx}
\begin{document}

\title{Poker as a Skill Game: Rational vs Irrational Behaviors}
\author{Marco Alberto Javarone}
\email{marcojavarone@gmail.com}
\affiliation{
Dept. of Mathematics and Computer Science, University of Cagliari, 09123 Cagliari, Italy
}

\date{\today}

\begin{abstract}

In many countries poker is one of the most popular card games. Although each variant of poker has its own rules, all involve the use of money  to make the challenge meaningful. 
Nowadays, in the collective consciousness, some variants of poker are referred to as games of skill, others as gambling. 
A poker table can be viewed as a psychology lab, where human behavior can be observed and quantified. This work provides a preliminary analysis of the role of rationality in poker games, using a stylized version of Texas Hold'em. 
In particular, we compare the performance of two different kinds of players, i.e., rational vs irrational players, during a poker tournament. 
Results show that these behaviors (i.e., rationality and irrationality) affect both the outcomes of challenges and the way poker should be classified.
\end{abstract}
\maketitle

\section{Introduction}\label{sec:introduction}%
Today modeling human behavior has become one of the most challenging topics in several scientific areas, e.g., psychology, social psychology, artificial intelligence and also statistical mechanics. 
In particular, sociophysics~\cite{galam01,galam02,galam03}, or social dynamics~\cite{loreto01}, address social systems, linguistics~\cite{javarone01,javarone02}, and also human behavior, adopting a strongly interdisciplinary approach, based mainly on the statistical mechanics framework.
Several aspects of social and human behavior have been modeled in the context of network theory~\cite{galam01,javarone03,tomassini01}, spanning from processes over social networks to analytical models based on spin glass dynamics~\cite{javarone04,weron01}.
We argue that poker games (hereinafter simply poker) could be considered as a natural environment and as a computational framework for studying human behavior~\cite{suzuki01}.
In particular, starting from this observation, we aim to introduce a well-known debate in the poker world: \textit{Is Poker a skill game?} This question is still considered an open problem, and the answer has a long list of implications~\cite{kelly01,cabot01}. 
Considering poker as gambling involves legal issues that vary from country to country and it also needs to be approached from a public health perspective (e.g., psychological dependence). In particular, the reason underlying these problems is associated with the use of money to play poker. Regardless of its classification (i.e., skill or gambling), poker is gaining prominence, to the point of being considered a sport or even a profession.
Poker is probably the most popular card game and much research in artificial intelligence has focused on devising algorithms to make machines capable of playing against humans ~\cite{dahl01,teofilo01}. 
From this perspective, codifying behavior such as the `bluffs' (later described), widely used both by experts and by amateurs, is not an easy task~\cite{seale01}. 
On the other hand computing a large number of solutions, in games such as chess, is now possible thanks to both artificial intelligence algorithms~\cite{norvig01}  and modern computational resources (e.g., high-frequency CPUs and parallel architectures).
There are several poker variants, e.g., Texas Hold'em, Omaha, Draw, etc., each having its own rules. However, they all follow a similar logic, i.e., a number of cards is distributed among players, who in turn have to decide whether to play or not, considering the possible combinations of their cards (called \textit{hand}) with those on the table. 
Firstly, the players place a  bet (money or chips), after having evaluated whether their hand could be the strongest. Hence, the use of money is fundamental to make the challenge meaningful, otherwise none of the players would have a reason to fold their hand. 
There are two main formats of playing poker, i.e., cash games or taking part in a tournament. In a tournament players pay an entry fee that goes into the prize pool plus a fee to play and receives a  certain amount in chips. Only the top players share the prize pool (usually money).
Of course the players' behavior is very important for increasing the chances of winning a poker challenge.  In particular, once players are dealt their hand, they have to decide whether or not to enter the pot (i.e., to play their hand) and if so, how much to bet. 
There are a number of betting rounds, where players try to evaluate the strength of their opponents hands (in terms of probability of winning the pot). During this phase, numerous variables are taken into account, e.g., the value of one's own hand, the size of the opponents' bet, the time opponents take to decide whether to bet or not, and so on. Hence, some kind of psychological analysis is performed by players during the challenge.
As mentioned above, players can bluff~\cite{guazzini01}, i.e., their behavior can deceive their opponents about the true strength of their hand. The derivation of the word poker suggests that it is mainly a game about bluffing and hence a psychological game.
Going back to the previous question, concerning the role of skills in poker, many conflicting considerations emerge and the matter is still open to debate as there is no definitive answer~\cite{hannum01}.
Here, we analyze this topic from a new perspective. In particular, we define a simple model, inspired by the Texas Hold'em variant, where agents play heads up challenges (i.e., agents faces just one opponent at a time) to win a tournament. 
In so doing, tournaments have a tree structure, where agents face each other, until only the last two players take part in the final challenge. 
Agents can behave rationally, i.e., they take decisions by evaluating their hands, or they can behave irrationally, i.e., they play randomly. Therefore, by comparing the performance of agents behaving in  opposite ways, we aim to provide a preliminary answer about the role of rationality in poker.
Although a comparison between rational strategies and random strategies may appear trivial, it is worth noting that, in some real scenarios such as stock markets, random trading strategies have been proven to be more useful than those based on `rational algorithms' (e.g., by using statistical patterns)~\cite{rapisarda01,rapisarda02}.
Therefore, investigations conducted in this work also represent a further attempt to evaluate the strength and the efficiency of random strategies in `partial information games'~\cite{dilemma01}.
Moreover, we also analyze these dynamics considering that rational agents can become irrational if, due to a series of bad hands, they lose a lot of money (or chips).
The main result of the simulations shows that, in the presence of a small number of rational agents, rationality is fundamental to success. Therefore, the classification of poker as a skill game does not depend on the game itself (i.e., its rules), but on the players behavior.
Furthermore, it is worth noting that when agents do change their behavior (i.e., from rational to irrational) this strongly affects the outcomes of the proposed model, further highlighting  the key role of human behavior in poker games.
The remainder of the paper is organized as follows: Section~\ref{sec:model} introduces the proposed model, for comparing the performance of rational and irrational agents. Section~\ref{sec:results} presents the results of the numerical simulations.  Lastly, Section~\ref{sec:conclusions} concludes the paper.
\section{The Model}\label{sec:model}
In the proposed model, we consider poker challenges which involve two opponents at a time. Furthermore, we use as reference a famous variation of poker called Texas Hold'em.
A brief description of Texas Hold'em is provided in the Appendix.
In so doing, we represent tournaments consisting of `heads up' challenges, where winning agents compete until such time as only two opponents contend for first place.
The agents’ behavior drives their actions during the challenges, i.e., they can be rational or irrational.
In order to evaluate their hands, rational agents use the Sklansky table~\cite{sklansky01}, a statistical reference for Texas Hold'em players.
On the other hand, irrational agents play randomly, hence they do not consider either the potential value of their hands or other factors.
The structure of each challenge, between two agents, represents a simplified version of a real scenario, as only one betting phase is performed (see Appendix for further details). Hence, rational agents act considering only the first two cards (i.e., their hand).
Moreover, we let \textit{blinds} (i.e., minimum bets posted by players) increase over time, as actually  happens in poker tournaments.
As in the real game, agents have to select among a set of possible actions: \textit{call}, \textit{bet}, \textit{raise}, and \textit{fold} --see Appendix. Generally, the first three actions involve the agent partaking in the pot, whereas the fourth represents withdrawal.
Comparing the performance of two agents, one rational versus one irrational, it is possible to compute the probability $P^w_r$ that rational agents will win a single challenge. 
Results of this preliminary analysis show that $P^w_r \in [0.8 - 0.85]$. 
Therefore, a rational agent wins about $W = 3$ consecutive challenges against an irrational one. It is worth observing that $W$ can be computed analytically once $P^w_r$ is known. 
In particular, considering $n$ consecutive challenges, the probability that the rational agent wins $n$ times is
\begin{equation}
P_r = \underbrace{ P^{w}_{r} \cdot P^{w}_{r} \cdot ... \cdot P^{w}_{r} }_{n} = (P^{w}_{r})^n
\end{equation} 
Therefore, as $n$ increases the value of $P_r$ decreases. As $W$ corresponds to the maximum value of $n$ such that $P_r \ge 0.5$, we get $n = 3$, then $W = 3$.
Furthermore, the value $W = 3$ has been confirmed also by performing $1000$ numerical simulations of challenges between two agents of different kinds (i.e., rational vs irrational).
In principle, considering the tree-like structure of heads up tournaments, a rational agent should be able to win when $N \le 2^W$. 
Hence, under this hypothesis, the proposed model yields the minimum theoretical density of rational agents (i.e., $\hat\rho_r$) to be able to consider poker as a `skill game'. 
In accordance with the value of $W$ derived above, the value of $\hat\rho_r$ can be computed as

\begin{equation}\label{eq:minimum_density}
\hat\rho_r = \frac{1}{2^W}
\end{equation}

\noindent recall that Eq.~\ref{eq:minimum_density} considers a tree-like structure of a heads up tournament. Since $W = 3$, we get $\hat\rho_r = 0.125$.
\subsection{Modeling variations of players' behavior}
Professional poker players, and generally expert players, know that adopting statistics-based strategies is only useful when they have a sufficient amount of chips (or money), otherwise an almost irrational strategy based on `all-in' actions is more incisive~\cite{sklansky01}. In particular, going `all-in' means that the player bets all her/his chips, in this case behaving almost irrationally as this move is driven not by the value of the hand but by the shortage of chips. Furthermore, it is worth noting that a rational player can also take this extreme action losing her/his patience after a consecutive series of bad hands, (i.e., she/he becomes irrational).
For the sake of simplicity, we model this phenomenon as follows: in the event a rational agent loses more than the half the amount of its starting capital (i.e., the initial amount of chips or money), while playing against an irrational agent, it can become irrational with the following probability

\begin{equation}\label{eq:transition}
p(R \to I) = 1 - C(t)/C(0)
\end{equation}

\noindent with $p(R \to I)$ transition probability from rational to irrational behavior, $C(t)$ amount of chips at time $t$, and $C(0)$ starting amount of chips. Recall that Eq.~\ref{eq:transition} is computed \textit{iff} at a given time $t$ the rational agent has less than $\frac{C(0)}{2}$ chips while playing against an irrational agent.
Note that usually, in real scenarios, players are able to recognize whether their opponents are completely irrational, as players have to show their hand after the last betting phase (see Appendix). Hence, actions disjointed by the value of the related hands are promptly identified and they result also quite different from unsuccessful attempts of bluff. Therefore the assumption underlying the proposed model (i.e., rational agents can become irrational only when they play against irrational opponents), is not too far from reality.
\section{Results}\label{sec:results}
\begin{figure}[!h]
\begin{center}
\includegraphics[scale=0.34]{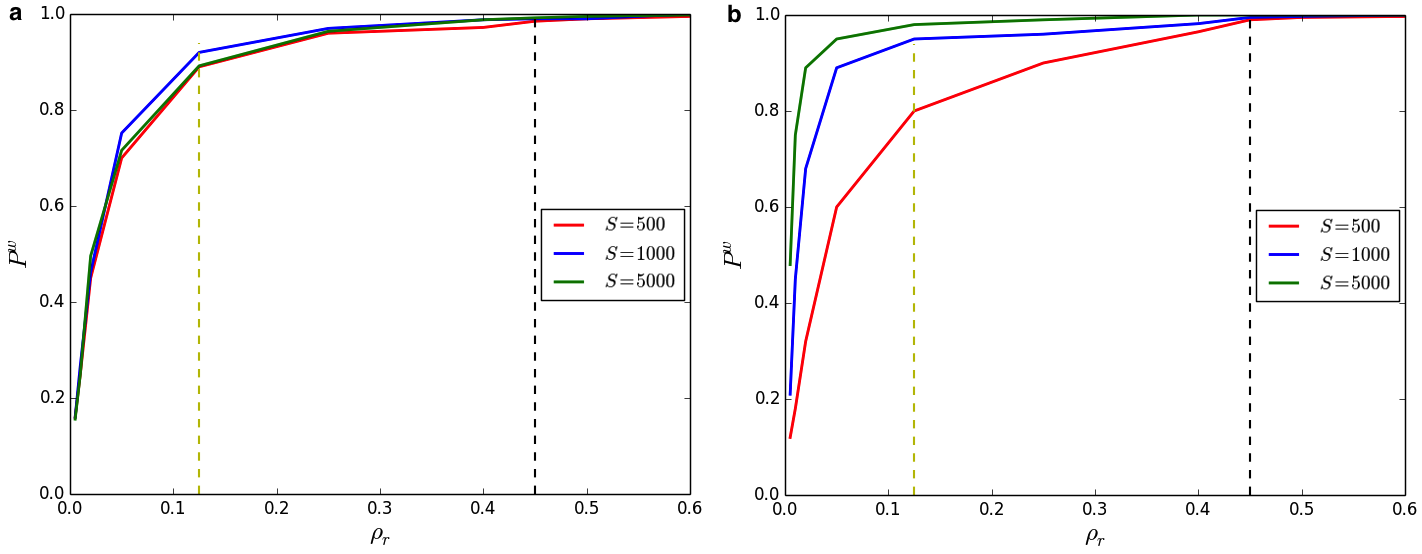}
\caption{Probability that a rational agent wins a tournament varying $\rho_r$ (i.e., density of rational agents). \textbf{a} Results achieved by the blinds sequence $\Sigma_1$. \textbf{b} Results achieved by the blinds sequence $\Sigma_2$. As indicated in the legends, each curve refers to a different starting stack $S$. Dotted lines indicate $\hat\rho_r$, i.e., the minimal theoretical density of rational agents to get always a rational winner: the yellow line refers to $2$ players, whereas the black one to tournaments. Results are averaged over $1000$ simulations.}
\label{fig:figure_1}
\end{center}
\end{figure}

In order to study the proposed model, we performed a large number of numerical simulations of tournaments with $N = 4096$ agents (i.e., players).
We consider three control parameters:
\begin{itemize}%
\item the density of rational agents $\rho_r$, \\with $\rho_r \in [0.005, 0.01, 0.02, 0.05, 0.125, 0.25, 0.4,0.45, 0.5, 0.6]$;
\item the starting stack $S$ assigned to agents for each single challenge (i.e., the initial amount of fiches), with $S \in [0.5k, 1k, 2k, 5k, 10k, 20k]$;
\item the sequence $\Sigma$ of \textit{blinds} (see Appendix).
\end{itemize}
In particular, we analyze two different sequences $\Sigma$: 
\begin{enumerate}
\item $\Sigma_1 = \{0.01k,0.02k,0.05k,0.1k,0.15k,0.3k,0.5k,0.8k,1k,1.5k,3k,5k,10k,20k\}$
\item $\Sigma_2 = 0.1k$
\end{enumerate}
The sequence $\Sigma_1$ is similar to those used in real tournaments where the \textit{blinds} gradually increase. On the other hand, $\Sigma_2$ does not represent a real scenario, but it is interesting to investigate the outcomes of the proposed model in the event \textit{blinds} do not vary over time.
Lastly, every $10$ time steps \textit{blinds} increase in accordance with the considered $\Sigma$.
We found that, as $\rho_r$ approaches $0.25$, the probability of a rational agent winning a tournament is higher than $0.95$ (for all values of $S$). In particular, as $S$ increases the minimum $\rho_r$, to get a rational winner, decreases ---see panel \textbf{a} of Figure~\ref{fig:figure_1}. 

As $\rho_r \to 0.4$, $P^w_{r} \to 1$ for all values of $S$.
Then, we analyzed how the probability $P^w_r$ varies in function of the starting stack $S$.
\begin{figure}[!h]
\begin{center}
\includegraphics[scale=0.6]{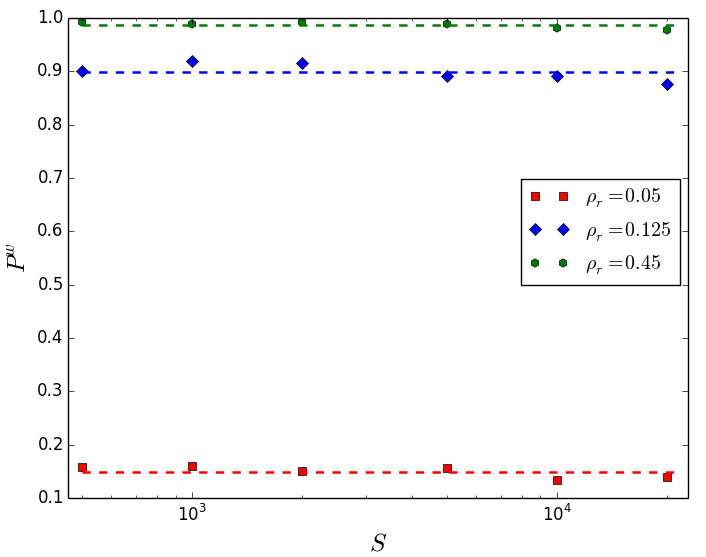}
\caption{Probability that a rational agent wins a tournament varying the starting stack $S$. As indicated in the legend, each symbol refers to a different density $\rho_r$ of rational agents, and dotted lines indicate the related average values. Results are averaged over $1000$ simulations.}
\label{fig:figure_2}
\end{center}
\end{figure}
Note that, as shown in Figure~\ref{fig:figure_2}, considering a different number of rational agents no relationship exists between the value of $S$ and $P^w$.
Lastly, we performed simulations considering constant \textit{blinds} over time, i.e., $\Sigma_2$.
Panel \textbf{b} of Figure~\ref{fig:figure_1} shows that increasing $S$, the minimum density $\rho_r$, to get a rational winner, decreases. In particular, the value of $P^{w}$ exceeds $0.8$ as $\rho_r = 0.01$. Therefore, by using constant \textit{blinds}, rational agents are at a great advantage.
\subsection{Changes of behavior}
\begin{figure}[!h]
\begin{center}
\includegraphics[scale=0.6]{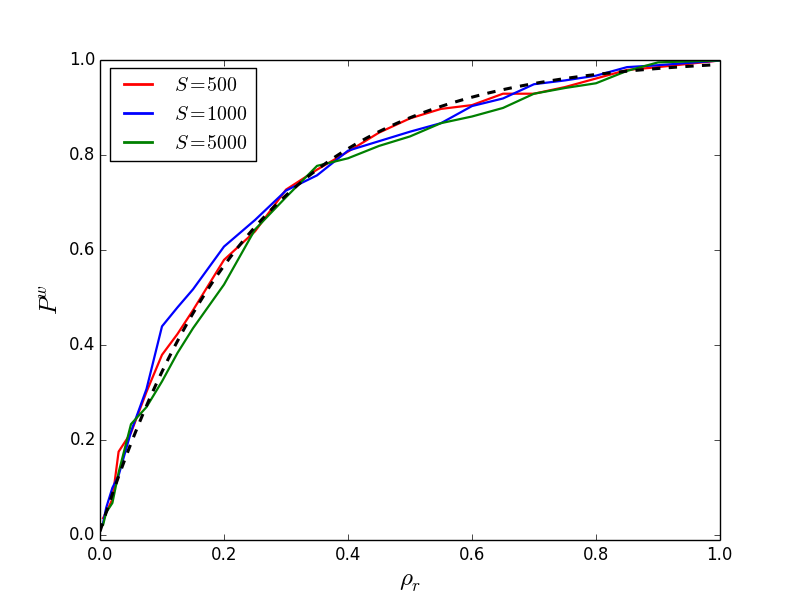}
\caption{Probability that a rational agent wins a tournament, by considering transitions $R \to I$ (i.e., from the rational to the irrational strategy), varying $\rho_r$ (i.e., density of rational agents). As indicated in the legend, each continuos curve refers to a different starting stack $S$, whereas the black dotted curve refers to computed fitting function. Results are averaged over $1000$ simulations.}
\label{fig:figure_3}
\end{center}
\end{figure}

\begin{figure}[!h]
\begin{center}
\includegraphics[scale=0.6]{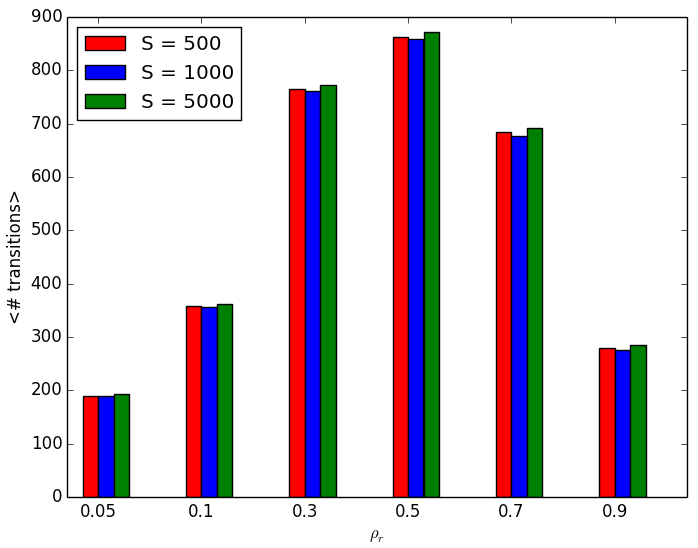}
\caption{Average number of transitions from rational to irrational behavior, varying $\rho_r$  (i.e., density of rational agents). Results are averaged over $1000$ simulations.}
\label{fig:figure_4}
\end{center}
\end{figure}

Here, we show the results obtained considering that rational agents, under the above described conditions (see Section~\ref{sec:model}), can become irrational. In this case, control parameters have the following values:
\begin{itemize}
\item $\rho_r \in [0.005, 0.01, 0.02, 0.03,0.05,0.075,0.1, 0.125,0.15,\\0.2, 0.25,0.3, 0.35, 0.4,0.45, 0.5,0.55, 0.6, 0.65,0.7, 0.75,0.8, 0.85, 0.9]$;
\item the starting stack $S$ assigned to agents for each single challenge (i.e., the initial amount of chips), with $S \in [0.5k, 1k, 2k, 5k, 10k, 20k]$;
\item the sequence of blinds is $\Sigma_1$ 
\end{itemize}
Figure~\ref{fig:figure_3} illustrate how the probability $P^w$ increases as $\rho_r$ increases.
We observe that the starting stack $S$ is irrelevant, since the value of $P^w$ is the same in all cases. Notably, we found that these curves (i.e., those related to $P^w$ varying $\rho_r$) can be well fitted by the function

\begin{equation}\label{eq:fitting}
P^w (\rho_r) = 1 - e^{-\lambda \cdot \rho_r}
\end{equation} 

\noindent with $\lambda \approx 4.1$, estimated by the non-linear least squares fitting method.
Therefore, the system behavior seems to have a non-linear relation with the amount of rational agents.
Since rational agents can change behavior over time, becoming irrational, we analyze the average number of transitions $R \to I$, i.e., transitions from rational to irrational behavior ---see Figure~\ref{fig:figure_4}.
We note that, in qualitative terms, the histogram of $<\# transitions>$ (i.e., average number of transitions $R \to I$) has a Gaussian-like form, peaked in $\rho_r = 0.5$.
As before, we did not find any correlation between the analyzed parameter (i.e., in this case $<\# transitions>$) and the initial stack $S$ provided to all agents.
\section{Discussion and Conclusions}\label{sec:conclusions}
\begin{figure}[!h]
\begin{center}
\includegraphics[scale=0.6]{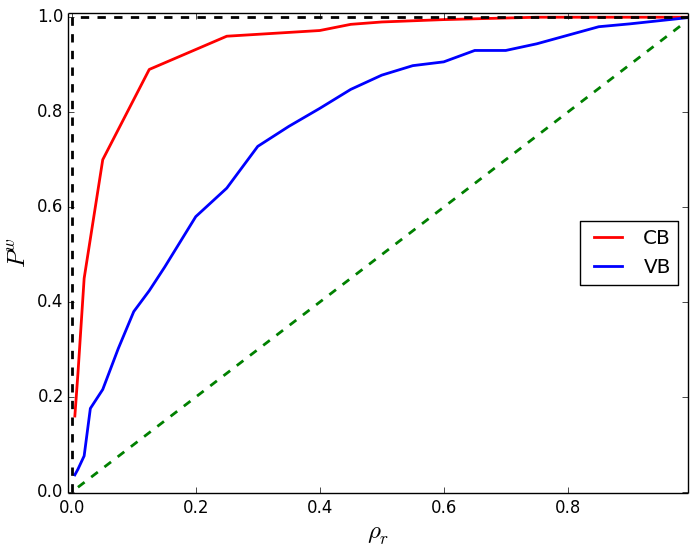}
\caption{Probability that a rational agent wins a tournament varying $\rho_r$  (i.e., density of rational agents). The red curve denotes agents with constant behavior (defined $CB$ in the legend), i.e., rational agents cannot become irrational. The blue curve denotes the variant of the proposed model where rational agents can vary their behavior (defined $VB$ in the legend) under opportune conditions.In both cases, the  curves refer to the results obtained with $S = 500$ (with $S$ starting stack). The black curve represents the step function, whereas the green one the simple linear function $P^w = \rho_r$. The $\epsilon$ on the $\rho_r$ axis corresponds to the minimum theoretical density of rational agents in a tournament (e.g., if $N = 1000$ then $\epsilon = \frac{1}{N} = 0.001$). Results are averaged over $1000$ simulations.}
\label{fig:figure_5}
\end{center}
\end{figure}

Our proposed model allows to compute the minimum theoretical density of rational agents to be able to consider poker as a `skill game'. 
We implemented two different versions of the model. The basic version considers agents with constant behavior i.e., rational agents always play  rationally, and irrational ones play always irrationally. 
The second version of the model allows rational agents to become irrational under opportune conditions. 
The main result of the numerical simulations, for both versions of the proposed model, suggests a preliminary answer to the above popular question `'Is Poker a skill game?: \textit{It depends on the players' behavior}''. 
This preliminary answer highlights the paramount importance of human behavior in poker games. In the first variant of the model, we compared two different sequences of \textit{blinds} and found that the structure of tournaments also plays an important role. 
For instance, in the event \textit{blinds} do not vary over time, rational agents are greatly advantaged. This latter phenomenon is widely known to poker players because as \textit{blinds} increases so too do the risks, therefore the `luck factor' can prevail over the players’ skills. 
On the other hand, we found that the initial stack ($S$) has little effect on these dynamics (see Figure~\ref{fig:figure_2}). This result warrants further analysis as usually, in real scenarios, the initial stack is deemed a significant parameter. 
The second variant of the model, provided further insights. Firstly, it is worth observing that the curve $P^w(\rho_r)$ can be well fitted by Eq.~\ref{eq:fitting}, therefore we identified a non-linear quantitive relation between the amount of rational players and the nature of poker. 
Furthermore, the maximum $\rho_r$ value for increasing the number of transitions ($R \to I$) is $\rho_r = 0.5$. This is not surprising as for lower values of $\rho_r$ there are fewer rational agents who can change their behavior. 
Instead, for values greater than $0.5$, there are too many rational players to trigger the transition (recall the transition can happen only when a rational agent plays against an irrational one). 
Here again, the parameter $S$ seems irrelevant. Lastly, we compared the two scenarios (i.e., the two variants of the model). 
As shown in Figure~\ref{fig:figure_5}, when poker is played with agents who do not change their behavior over time, it resembles a skill game more closely, whereas when their behavior changes, then the nature of the game tends more to gambling. 
In Figure~\ref{fig:figure_5}, we identify two theoretical curves to discriminate between skill games and gambling. 
A step function (whose minimum value $\epsilon$ corresponds to the minimum theoretical density of rational agents in a tournament) identifies pure skill games, inasmuch as in the presence of only one rational agent, it wins the tournament (e.g., Chess). 
On the other hand, the linear function $P^w(\rho_r) = \rho_r$ represents gambling (e.g., if $\rho_r = 0.3$ then the $P^w = 0.3$). 
Comparing the distance of a curve, obtained for a particular game, from the two theoretical ones described above (i.e., the skill and the gambling curve) may also be useful for defining the nature of other games, hence it could prove a valuable quantitative evaluation tool. 
In brief, we found that, even if only a few players play rationally, rationality becomes a fundamental ingredient to success. As a consequence, poker can be considered a skill game only when its players behave rationally. 
Therefore, in these games human behavior is more important than the rules themselves for understanding the nature of the game. To conclude, although it would be extremely interesting to compare the outcomes of the proposed model with real data, this is not possible as no similar datasets exist. 
For single challenges, it may be possible to evaluate whether a player is adopting mainly a random strategy or a rational strategy. 
Moreover, in real scenarios, a completely random strategy is adopted by very few players. 
It is worth recalling that, in this context, we focused the attention on evaluating whether a comparison between a random vs a rational strategy may provide a preliminary answer as to the nature of Poker. 
As for future work, we aim to further analyze the proposed model in order to make it as realistic as possible and to represent other kinds of human behavior observable during poker challenges. 

\section*{Appendix}\label{sec:appendix}
Here, we briefly illustrate the basic rules of one of the most well-known variants of poker, namely Texas Hold'em~\cite{sklansky01}. 
Let us consider two players playing a heads up challenge. 
At the beginning of the challenge, both players have an equal stack of chips (e.g., 10000\$). The winner is the agent who wins all the chips from its opponent. 
Each round is won by the agent with the best combination of five cards, computed from the seven available (a further explanation is given below). 
Each round (say time step) of the challenge comprises four phases, where agents have to decide whether to place a bet or fold, depending on their cards and on the strategy adopted by their opponent. 
They are dealt two cards (their hand) and then place a small amount of chips in the pot (the prize for the single round composed of all the bets), then the round begins. 
During each phase, agents have to select among the following actions: \textit{call}, \textit{fold}, \textit{bet}, and \textit{raise}. To \textit{call} means to put in the pot the same amount of chips as their opponent. 
To \textit{fold} means to drop out of the hand, in which case the opponent wins the whole pot and the round ends. To \textit{bet} means to increase the pot using some of the chips. 
In this case, the opponent has to decide whether to accept to post the same amount of chips or not. Lastly, to \textit{raise} means to increase the bet posted by the opponent. These betting phases alternate during each single round. 
During the first phase (i.e., the \textit{pre-flop}) agents only have their own ($2$) cards. Then in the second phase, called \textit{flop}, three cards are dealt onto the table. Now, in the \textit{turn} phase, a further card is shown. Lastly, during the \textit{river} phase, the last card is dealt. 
As discussed before, the winner is the agent with the best combination of five cards, using the cards in their hand and those on the table (called \textit{community cards} that can be used by both players). In the event the best combination is composed of the \textit{community cards}, both agents win, and each one receives half the pot. 
During the above described phases, the agents decide the best of the above actions to take in order to win the pot. One important aspect to highlight when describing a poker challenge, concerns time. 
Every time a player makes a \textit{bet}, there exists a minimum number of chips that can be posted called \textit{blind}. The size of the \textit{blind} increases over time at regular time intervals (e.g., $15$ minutes, $20$ minutes, etc.). 
Therefore, first rounds are usually characterized by small pots, whereas the last rounds are much richer. Finally, the challenge ends when one agent wins all the chips from its opponent. 
Although the basic rules of Texas Hold'em are fairly simple, even with the aid of a device for computing winning probabilities (of each hand), it is not so easy to win a challenge. 
In particular, many kinds of psychological behavior drive players during the selection of their actions. Particularly, as discussed before, the opportunity to perform bluffs makes this game non-trivial.

\end{document}